# Hiding Malicious Content in PDF Documents

## Dan-Sabin Popescu

*Information Technology Security*
*The Military Technical Academy*
*George Coşbuc 81-83, Bucharest 5, 050141, ROMANIA*
*sabin.popescu@yahoo.com*
http://www.mta.ro

**Abstract**: This paper is a proof-of-concept demonstration for a specific digital signatures vulnerability that shows the ineffectiveness of the WYSIWYS (What You See Is What You Sign) concept. The algorithm is fairly simple: the attacker generates a polymorphic file that has two different types of content (text, as a PDF document for example, and image: TIFF – two of the most widely used file formats). When the victim signs the dual content file, he/ she only sees a PDF document and is unaware of the hidden content inside the file. After obtaining the legally signed document from the victim, the attacker simply has to change the extension to the other file format. This will not invalidate the digital signature, as no bits were altered. The destructive potential of the attack is considerable, as the Portable Document Format (PDF) is widely used in e-government and in e-business contexts.
**Key-Words**: Digital signatures, vulnerabilities, PDF

## 1. Introduction

The digital signature, as defined by Diffie and Hellman [1], is a widespread application of asymmetric key cryptography, whose purpose is to ensure the integrity of the signed documents and to guarantee the identity of the signer. In many countries, digital signatures can legally substitute handwritten signatures [2] and are widely used in e-business and e-government activities.

To digitally sign a document, one must first generate a hash of the original file and encrypt this digest using the private key of an asymmetric algorithm (RSA being one of the most popular). Any tampering will automatically invalidate the signature, as the hash value calculated during decryption will not match the original.

However, digital signatures aren't perfect, as they allow the content of a file to be displayed dynamically [3], depending on various instructions included in the file (PDF files can incorporate JavaScript sequences, for example). This function is useful if you would like to have a quotation document updated automatically with the latest exchange rates. Dynamic content does not invalidate a digital signature and gives attackers a whole new area to explore (and to exploit). To get around this, the WYSIWYS (What You See Is What You Sign) concept was introduced. In short, because the binary and hex interpretations of a document are incomprehensible to most people, the file is converted into a static image (like BMP or TIFF) before being signed.

By exploiting the file structures of various text and image formats, an attacker can obtain a legally signed document, but whose content differs from the one that the signer originally approved.

## 2. Preliminary Info

Thus, what the user sees on the screen is actually not what he signs. As stated above, the mechanism is straight forward: thettacker generates a dual file that includes both a PDF document and a TIFF image. When the victim signs the polymorphic file, he/ she only sees a PDF file and is unaware of the hidden content. After obtaining the legally signed document from the victim, the attacker simply has to change the extension to the other file format. This will not invalidate the digital signature, as no bits of the actual file were altered.

This method is known as the Dali Attack, named after the famous painter [4], [5]. The first demonstration of the attack was based on BMP and HTML files. Assume the following scenario: the CEO of a company wants to grant 100,000 Euros to the financial department, so he asks the CFO to write up a document. To ensure there are no hidden macros or scripts, the CEO demands a scanned BMP copy of the document. The CFO, who is the attacker in this scenario, wants to gain more funds, so he inserts a hidden HTML code inside the image file. This code is actually the same document, where the 100,000 Euros amount is changed to 1 million Euros.

When the CEO digitally signs the Grant.bmp file, he is unaware of the hidden code behind the document. His smartcard device or software-based cryptography application generates the Grant.bmp.p7m file (PKCS#7). The file then goes back to the CFO, who changes the extension to Grant.htm.p7m. Because the digital signature verification is done solely by comparing the bits that make up the PKCS container (of which none references the filename or extension), the signature will still be valid.

If one would open the file with an image viewer, one would still get the original document, with the approved 100,000 Euros amount. However, because the extension was changed to HTML, the operating system will automatically use a web browser to open the file, thus displaying the modified text.

The biggest problem of this kind of attack is the use of HTML files, that aren't usually encountered in a typical corporate document workflow. Plus, an HTML text file that is over 2 MB will surely raise suspicions.

The Dali Attack can be improved by using TIFF and PDF files that have a very flexible structure. For example, TIFF files allow you to store image parameters (resolution, dimensions etc.) anywhere inside the file. On the other hand, PDF documents are read from the end of the file towards the beginning and the header can be placed anywhere within the first 1,024 bytes of the file.

If executed correctly, the attacker can generate an almost undetectable polymorphic file, which can be used for fraud in practically any environment that relies on PDF for the internal document workflow.

Before describing the attack in detail, we must first take a look at the basics of the PDF and TIFF file structures.

## 2.1. The PDF File Structure

PDF is a platform independent standard developed by Adobe Systems for electronic documents exchange. The main sections of the PDF format are [7]:

▪ Header – identifies the PDF version (for compatibility reasons); it is usually defined as %PDF-1.X[EOL], where X is in range {0, 7 > the latest version of Acrobat, 9.0}, and EOL is the End-of-Line marker, usually CR (Carriage Return, 0D in hex), LF (Line Feed, 0A in hex) or both. It can occur anywhere within the first 1,024 bytes of the file;

▪ Body – the visual components of the file (text, images, fonts, pages layout, objects etc.);

▪ Xref (Cross-reference Table) – pointers and other information about the various embedded objects; it allows Adobe Reader to find objects anywhere within document, by searching for the corresponding offset. Thus, the PDF viewer doesn't have to scan the whole file to find an object.

▪ Trailer – specifies the location of the Xref Table and of other objects.

▪ The PDF format is designed to be read from the end, in order to quickly find the Xref Table. The last line of the document must contain the %%EOF marker (End-of-File).

Figure 1 illustrates the basic file structure of a PDF file.

**Header**

```
        % PDF -1.4 .
[hex]25 504446 2D31E34 0D
         Ver    EOL
               (CR)
```

**Body**

```
... <</Length 66/Filter/
FlateDecode/I 8&/L 70/S 38>> ...

... <</Font<</TT2 10 0 R>> ...

... obj.<</Subtype/TrueType/
FontDescriptor 12 0 R/LastChar
117/Widths[250 0 0 0 0 0 0 0 0
0 0 0 0 0 0 250 0 500 500
0 0 667 0 0 0 0 0 0 0 0 0
0 0 556 0 0 556 0 0 0 0
500 444 0 0 0 444 0 444 0 778
500 500 0 0 333 0 278 500]/
BaseFont/TimesNewRomanPSMT/
FirstChar 32/Encoding/
WinAnsiEncoding/Type/Font>>
.endobj.11 0 ...
```

**Cross-reference Table**

```
xref..
0  6.. Subsection x 6
0000000000 65535 f..
30303030303030303035353533333520 640304
   Object #0    Space    EOL
                        (CR+LF)
0000001831 00000 n..
0000001665 00000 n..
0000001889 00000 n..
0000001940 00000 n..
0000005609 00000 n..
              f = free; n = in-use
```

**Trailer**

```
trailer..
<</Size 6>>..      # Total Xref
startxref..
116..              Offset Xref
%%EOF..             End-of-File
```

*Figure 1: Outline of the PDF format structure*

## 2.2 The TIFF File Structure

TIFF, a format widely used for manipulating high resolution images, was developed by Aldus, a company acquired by Adobe Systems in 1994. TIFF has a very flexible structure, which envelops all the image data in structures called IFDs (Image File Directories). IFDs are two-dimensional arrays that specify image resolution, compression, the total number of colours used etc., as well as the pointers that define the offsets of these parameters.

Because IFDs can be placed anywhere within the TIFF file, the document must contain a pointer to the first IFD. This pointer is placed inside the eight byte header of the TIFF file.

The first two bytes identify the TIFF format and byte order (4949 [hex] or "II" [ASCII] for little-endian and 4D4D [hex] or "MM" [ASCII] for big-endian). The next two bytes are the so-called "magic number" (002A [hex] or 42 [decimal]), also used to identify the TIFF format. The last four bytes of the header are the offset (starting address) of the first IFD. The TIFF specifications [6] do not specify constraints in regard to this offset, which means it can even be placed at the end of the file. This is a huge advantage for the attacker – he has the possibility to insert an arbitrary code (in this case a PDF document) immediately after the header.

An IFD structure starts with a two byte sequence that specifies the total number of directories (components). The last four bytes define the offset of the next IFD (if it exists). In between, there are multiple 12-byte one-dimensional arrays that define all of the image parameters and are structured as follows:

- Bytes 0-1: Tag – the identifier;
- Bytes 2-3: Type (Byte/0001h, unsigned int, 8b; ASCII/0002h,
- 7b+NUL; Short/0003h, unsigned int, 16b; Long/0004h, unsigned int, 32b; Rational/0005h, 2xLong); TIFF Revision 6.0 includes seven types (signed versions of the above mentioned types, plus Float/0011h and Double/0012h);
- Bytes 4-7: Count (Length) – the total number of values;
- Bytes 8-11: Value/ Offset – specifies the address (byte-wise) where the value of the field is stored at; the field contains the actual value <u>if and only if</u> it is smaller than 4 bytes.

Black and white images can be defined using just the following IFD subfields:

- PhotometricInterpretation/0106h;
- Compression/0103h;
- ImageLength/0101h;
- ImageWidth/0100h;
- ResolutionUnit/0128h;
- XResolution/011Ah;
- YResolution/011Bh;
- RowsPerStrip*/0116h;
- StripOffsets*/0111h;
- StripByteCounts*/0117h;

* The TIFF image is split into strips, which make it easier to edit the image and also optimizes the input/ output buffer. Thus, similarly to PDF files, the image viewer does not have to scan the whole file to find a specific parameter.

Greyscale images have an additional field called BitsPerSample/0102h, while full RGB (Red, Green, Blue) images also use the SamplesPerPixel/0115h field.

The general structure of an IFD and that of a TIFF file are described in Figures 2 and 3.

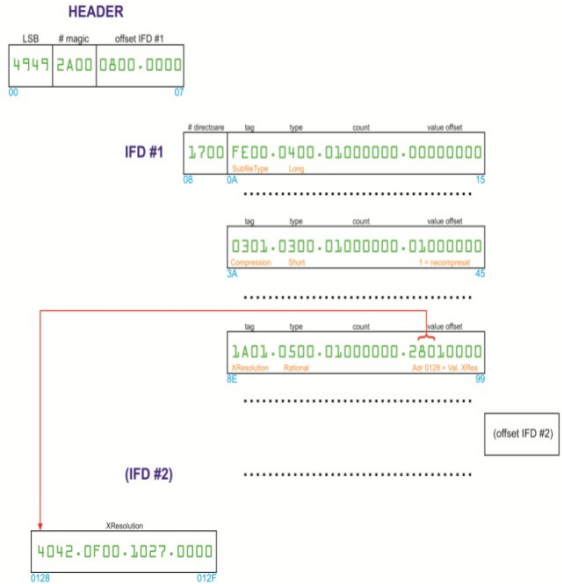

*Figure 2. The header and first IDF of a TIFF file*

| Offset (hex) | Description | | Value (numeric values are expressed in hexadecimal notation) |
|---|---|---|---|
| **Header:** | | | |
| 0000 | Byte Order | | 4D4D |
| 0002 | 42 | | 002A |
| 0004 | 1st IFD offset | | 00000014 |
| **IFD:** | | | |
| 0014 | Number of Directory Entries | | 000C |
| 0016 | NewSubfileType | 00FE 0004 | 00000001 00000000 |
| 0022 | ImageWidth | 0100 0004 | 00000001 000007D0 |
| 002E | ImageLength | 0101 0004 | 00000001 00000BB8 |
| 003A | Compression | 0103 0003 | 00000001 8005 0000 |
| 0046 | PhotometricInterpretation | 0106 0003 | 00000001 0001 0000 |
| 0052 | StripOffsets | 0111 0004 | 00000BC 000000B6 |
| 005E | RowsPerStrip | 0116 0004 | 00000001 00000010 |
| 006A | StripByteCounts | 0117 0003 | 00000BC 000003A6 |
| 0076 | XResolution | 011A 0005 | 00000001 000006F6 |
| 0082 | YResolution | 011B 0005 | 00000001 000006FE |
| 008E | Software | 0131 0002 | 0000000E 000006A6 |
| 009A | DateTime | 0132 0002 | 00000014 000006B6 |
| 00A6 | Next IFD offset | | 00000000 |
| **Values longer than 4 bytes:** | | | |
| 00B6 | StripOffsets | | Offset0, Offset1, ... Offset187 |
| 03A6 | StripByteCounts | | Count0, Count1, ... Count187 |
| 0696 | XResolution | | 000012C 00000001 |
| 069E | YResolution | | 000012C 00000001 |
| 06A6 | Software | | "PageMaker 4.0" |
| 06B6 | DateTime | | "1988:02:18 13:59:59" |
| **Image Data:** | | | |
| 00000700 | | | Compressed data for strip 10 |
| xxxxxxxx | | | Compressed data for strip 179 |
| xxxxxxxx | | | Compressed data for strip 53 |
| xxxxxxxx | | | Compressed data for strip 160 |

*Figure 3. The hex outline of a TIFF file*

*Note: Fields like Date/ Time, Software, Artist, ICCProfile are optional and most image viewers and editors are designed to ignore them if the data is non-interpretable.*

## 3. Embedding the Malicious Content

The goal of the attack is to obtain a dual PDF/ TIFF file that shows the two different types of content by changing the file extension. The flexible structure of the TIFF format allows the insertion of arbitrary code without the risk of corrupting the file. However, when any additional number of bytes are added, to pointers are shifted, and therefore it is necessary to manually adjust the offsets of the header and of the IFD components. PDF is also a good choice for this type of attack, because the header can be placed anywhere within the first 1,024 bytes, which means that it can be preceded by an eight byte sequence that is actually a TIFF header.

We will consider the scenario described in Chapter 2: the CFO of a company wants to gain access to funds of 1 million Euros, rather than the approved 100,000 Euros.

The first step is to generate the two files – Contract.pdf and Contract.tif (containing the modified amount). Then, we copy the whole content of the PDF file after the first eight bytes of the TIFF image. At the end of the new polymorphic file, we add the trailer (last few bytes) of the PDF document, in order to preserve compatibility with Adobe Reader. All the operations can be done using any hex editor. I opted for Hex Workshop 4.2. At this point, the polymorphic file can be opened by Adobe Reader, but image viewers return an error because the offsets are wrong.

Next, we must modify the header, which contains the offset of the first IFD, respectively the last four bytes of the 4949.2A00.0800.0000 sequence. In the original file, the 8h address corresponds to the 1700.FE00.0400.0100 sequence. After inserting the PDF file, all values are shifted with the byte-value equivalent of the PDF document (6,009 bytes). To determine the new IFD position, we must add this value to the original position. In hex, 6,009 is 1779h. Thus, 1779h + 0008h = 1781h.

Another method to determine the new offset (which is the method used in the video file on the CD submitted with this paper), is to search for the 1700.FE00.0400.0100 sequence in the polymorphic file – which will point to the 1781h location, as below:

```
00001750  206E 0D0A 7472 6169 6C65 720D 0A3C
00001760  5369 7A65 2036 3E3E 0D0A 7374 6172
00001770  7265 660D 0A31 3136 0D0A 2525 454F
00001780  0A37 00FE 0004 0001 0000 0000 0000
00001790  0103 0001 0000 0052 0300 0001 0103
000017A0  0000 004C 0400 0002 0103 0003 0000
000017B0  1800 0003 0103 0001 0000 0001 0000
```

*Figure 4. The new location of the IFD Now, all we have to do is change the header to 4949.2A00.8117.0000 (1781h becomes 8171h in big-endian).*

The same process has to be repeated for most of the IFD parameters. To identify them more easily, I used an application called AsTiffTagViewer 2.0.

For the polymorphic file to be nearly identical to the original TIFF file, the following fields must be changed: BitsPerSample, XResolution, YResolution and StripOffsets. For a more convincing result, one should also adjust the other readable fields (username, the software used to generate the file, the date and time of creation etc.).

Table 1, available at the end of this paper, illustrates all of the changes that must be made to the polymorphic file.

After this, the finished polymorphic file Contract.pdf.tif, which will be renamed Contract.pdf to fool the victim, can be opened without error by Adobe Reader or Foxit Reader. Furthermore, when analyzed by the Preflight tool in Acrobat Professional, the polymorphic file appears to be a standard PDF file, with no syntax errors.

Accessed with any image viewer, the polymorphic file will display the malicious content. The only tool that detects any irregularities with the file is Adobe Photoshop, which simply states the image has data that cannot be read. After this, the attacker sends the polymorphic file to his victim, who signs it using a smartcard device or a software solution capable of handling digital certificates.

The resulting Contract.pdf.pkcs7 file will pass signature verification, because all the malicious modifications were made before signing. When decoded with signature verification software, the user will view the original PDF document, with the 100,000 Euros amount. To complete the attack, the attacker must change the filename extension from Contract.pdf.pcks7 to Contract.tif.pkcs7. The digital signature verification process will once again pass (see Figure 5 below), because the bits that make up the PKCS#7 message contain no information about the extension of the file. Any operating system will then interpret the polymorphic file by its new extension (TIFF) and open it with an image viewer, displaying the modified 1 million Euros amount.

This type of attack also works for password protected PDF documents (RC4 or AES), as well as PDF/A documents (used for long term archiving), PDF/E documents (used for engineering workflows) and PDF/X documents (used in the desktop publishing and prepress industry).

It is also worth mentioning that the Dali Attack is quite difficult to detect because it does not cause direct damage to the victim (like credit card fraud for example). The attacker gains an advantage that he can exploit sometime in the future.

# 4. Methods of Detection

I have identified seven different solutions (freeware or commercial) that could help in identifying a supposed polymorphic file. Some methods are for the tech savvy persons that are willing to open the suspicious file with a hex editor (a direct method of identifying any tampering) or with Adobe Photoshop (that will detect any errors in the TIFF format).

In the case of PDF/A files, the Preflight tool in Adobe Acrobat Professional will detect syntax errors if the file is put through the PDF/A-1b standard compliancy test. This standard states that the PDF header must begin with %PDF-1.X, and cannot start with an arbitrary code.

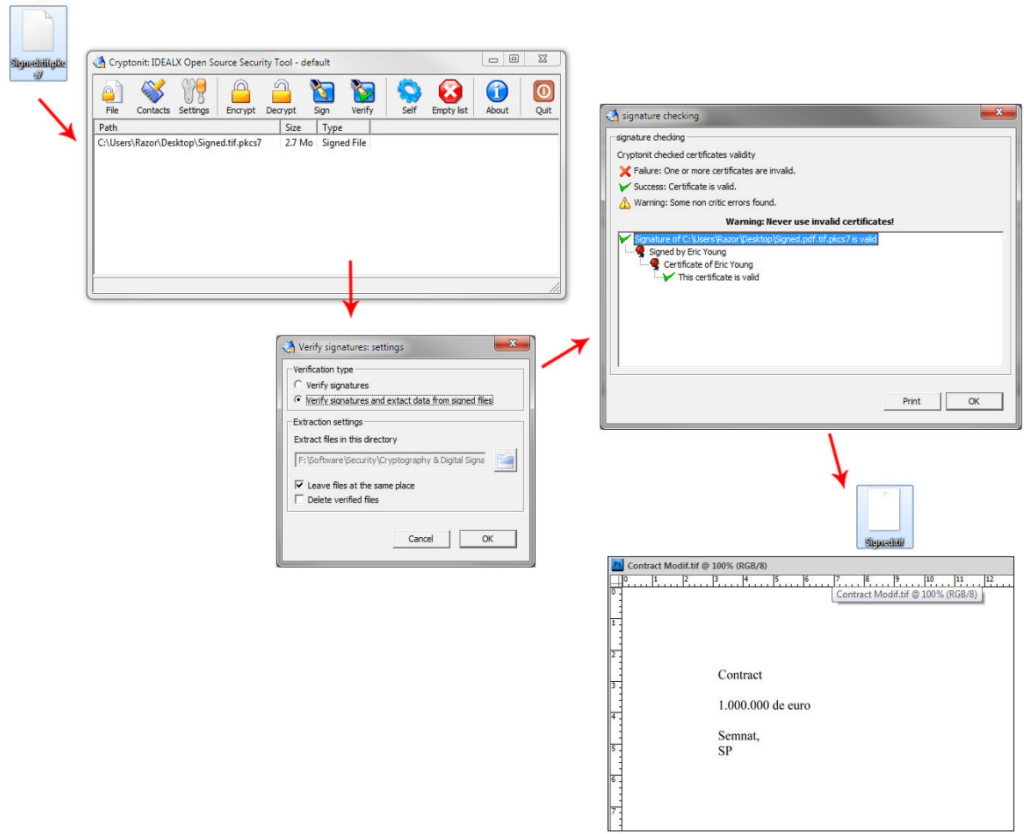

*Figure 5: Signature verification of the polymorphic file (renamed to Contract.tif.pkcs7)*

On the other hand, if a user already owns Acrobat Professional, then he has a sure method of disarming any kind of Dali Attack, because the proprietary Adobe signature software rewrites the whole document, eliminating any prior modifications [8]. Another complex method would be to include the filename or corresponding MIME ContentType in the PKCS#7 container. Thus, the signature verification software would detect when the attacker tries to change the extension.

One could also develop a heuristic application that could search for patterns specific to the Dali Attack. Any inexperienced user could then detect an attack.

However, because developing such an application is fairly complex (requiring implementing CMS and PKCS#7 libraries in IDEs like .NET), I have developed a simple batch program that can detect the attack and then display the TIFF image hidden inside a PDF document.

This batch is intended to be run before signing a suspicious file. The underlying principle is simple: using a tool from the ImageMagick 6.6 suite, we search for TIFF image specific parameters inside a PDF document (like dimensions, resolution, color depth etc.). If they are present, then the file might be polymorphic. The batch then duplicates the file, renames it to TIFF and opens the default image viewer. The Identify tool from ImageMagick can also detect PDF specific fields (like format: application/pdf and pdf:Producer:Acrobat Distiller 8.1.0), meaning it can also be used to detect images that have a PDF document hidden inside.

The batch file (compiled for both x32 and x64 platforms) is included on the CD submitted with this paper.

Figure 6 below shows the output of the batch program, that has detected TIFF parameters inside a PDF file.

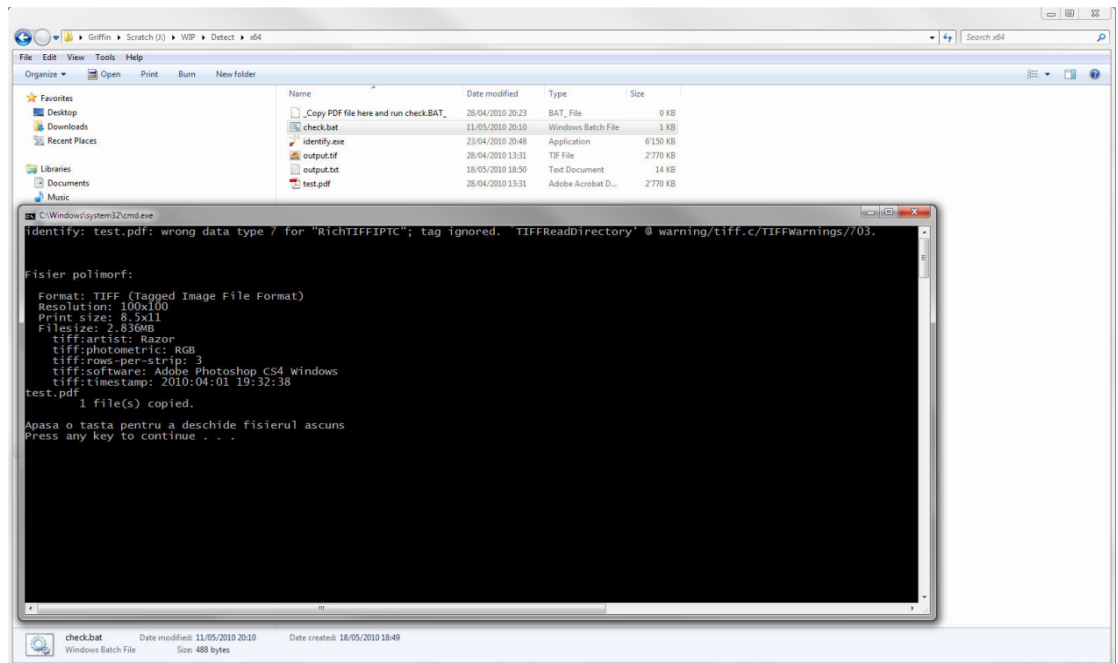

*Figure 6. The output of the batch program designed to detect polymorphic file*

## 5. Conclusions

This paper and the video attached to it describe in detail a new digital signatures exploit that allows an attacker to trick his victim into signing a document he/ she did not approve. The attack is not based on embedding dynamic content (like macros in Microsoft Word) or accessing external components (such as fonts in a PDF file), but rather on creating a polymorphic file with two types of content
– TIFF and PDF – one being the original document and the other, the modified malicious copy. Thus, the victim is unaware of signing another document hidden behind the one displayed on screen. The destructive potential of the attack is considerable, as both PDF and TIFF are widely used in e-government activities, respectively in the corporate environment.
I have also listed the main methods of detecting or disarming this type of attack, including an application I have developed myself.

**Table 1**: The changes needed to be done to the original TIFF image to envelop the PDF document